\documentclass[prl, aps, twocolumn, superscriptaddress]{revtex4-1}
\usepackage{graphicx,graphics,psfrag,amsmath,calc,mathtools}
\usepackage{amsfonts}
\usepackage{epsfig, bm}
\usepackage{color,comment}
\usepackage{cancel}
\usepackage{float}

\usepackage{hyperref}
\hypersetup{
  colorlinks   = true, 
  urlcolor     = blue, 
  linkcolor    = blue, 
  citecolor   = red 
}

\usepackage{graphicx}
\usepackage{physics,bm,ulem}
\usepackage{color}
\usepackage{lineno}

\usepackage{color}

\begin{document}

\title{Quantum state discrimination in a $\mathcal{PT}$-symmetric system of a single trapped ion}

\author{Chenhao Zhu$^{\#}$}
\affiliation{Department of Physics and Beijing Key Laboratory of Opto-electronic Functional Materials and Micro-nano Devices, Renmin University of China, Beijing 100872, China}
\author{Tingting Shi$^{\#}$}
\affiliation{Department of Physics and Beijing Key Laboratory of Opto-electronic Functional Materials and Micro-nano Devices, Renmin University of China, Beijing 100872, China}
\author{Liangyu Ding}
\affiliation{Beijing Academy of Quantum Information Sciences, Beijing 100093, China}
\author{Zhiyue Zheng}
\affiliation{Beijing Academy of Quantum Information Sciences, Beijing 100093, China}
\author{Xiang Zhang}
\thanks{siang.zhang@ruc.edu.cn}
\affiliation{Department of Physics and Beijing Key Laboratory of Opto-electronic Functional Materials and Micro-nano Devices, Renmin University of China, Beijing 100872, China}
\affiliation{Beijing Academy of Quantum Information Sciences, Beijing 100093, China}
\affiliation{Key Laboratory of Quantum State Construction and Manipulation (Ministry of Education), Renmin University of China, Beijing 100872,China}
\author{Wei Zhang}
\thanks{wzhangl@ruc.edu.cn}
\affiliation{Department of Physics and Beijing Key Laboratory of Opto-electronic Functional Materials and Micro-nano Devices, Renmin University of China, Beijing 100872, China}
\affiliation{Beijing Academy of Quantum Information Sciences, Beijing 100093, China}
\affiliation{Key Laboratory of Quantum State Construction and Manipulation (Ministry of Education), Renmin University of China, Beijing 100872,China}

\date{\today}

\begin{abstract}
We experimentally demonstrate an unambiguous quantum state discrimination of two qubit states under a non-Hermitian Hamiltonian with parity-time-reversal ($\mathcal{PT}$) symmetry in a single trapped $^{40}$Ca$^+$ ion. We show that any two non-orthogonal states can become orthogonal subjected to time evolution of a $\mathcal{PT}$-symmetric Hamiltonian in both the $\mathcal{PT}$-symmetry preserving and broken regimes, thus can be discriminated deterministically. For a given pair of candidate states, we show that the parameters of the Hamiltonian must be confined in a proper range, within which there exists an optimal choice to realize quantum brachistochrone for the fastest orthogonalization. Besides, we provide a clear geometric picture and some analytic results to understand the main conclusions. Our work shows a promising application of non-Hermitian physics in quantum information processing.
\end{abstract}
\maketitle 

{\it {Introduction.}}-- Quantum state discrimination (QSD) is of great importance in a wide range of applications of quantum computation and communication~\cite{barnett-1997-1,chefles-2000,barnett-2001,bergou-2004,chefles-2004,croke-2009,bergou-2010,bae-2015}. In many practical quantum tasks, information encoded in quantum states needs to be read out and discriminated among a set of candidates~\cite{nelson-2000}. In a Hermitian system, if the candidate states are mutually orthogonal, the decoder can work perfectly by measurements of observables along their directions. However, when the candidates are non-orthogonal, QSD can only be realized with some error. To enhance the confidence level in state discrimination, several optimal measurement schemes~\cite{croke-2006,helstrom-1976,ivanovic-1987} with respect to various criteria have been constructed, including the minimum-error discrimination~\cite{helstrom-1976,barnett-1997,clarke-2001} and the unambiguous state discrimination~\cite{ivanovic-1987,dieks-1988,peres-1988,huttner-1996,clarke-2001-2}.

Recent studies on non-Hermitian systems with parity-time-reversal ($\mathcal{PT}$) symmetry~\cite{bender-1998} reveal many novel features and potential applications, such as information retrieval~\cite{ueda-2017,ding-2022}, quantum brachistochrone~\cite{bender-2007,samsonov-2008,long-2013}, super quantum temporal correlations~\cite{wu-2023, quinn-2023}, and enhancement of sensitivity~\cite{yang-2017,hodaei-2017}. $\mathcal{PT}$-symmetric systems can be well prepared and manipulated with high precision in both classic~\cite{peng-2014,ruter-2010,kottos-2011} and quantum~\cite{murch-2019,xue-2017,luo-2019,chen-2021,ding-2021,ding-2022,cao-2023} platforms, confirming many intriguing predictions and stimulating further investigations. 
For the task of QSD, it is proposed that with the help of a $\mathcal{PT}$-symmetric Hamiltonian one can orthogonalize two non-orthogonal quantum states and discriminate them deterministically in a single measurement at the cost of a null probability~\cite{bender-2013}. This so-called $\mathcal{PT}$-symmetric QSD approach works when discriminating a given quantum state from two possible non-orthogonal candidates known {\it a priori}. The null probability arises from the similarity transformation from the initial Hilbert space spanned by the non-orthogonal states to the final Hilbert space determined by the eigenstates of the $\mathcal{PT}$-symmetric Hamiltonian, which is analogous to the mechanism of unambiguous state discrimination. The $\mathcal{PT}$-symmetric QSD only requires a single phenomenological parameter of decay rate to describe the bath, hence can be applied to more general scenarios where detailed knowledge of the auxiliary space is lack.
Recently, the $\mathcal{PT}$-symmetric QSD was generalized to the case of three candidate states~\cite{chang-2021}, and demonstrated using a lossy linear optical setup~\cite{nori-2022}. Nevertheless, studies in this field are restrained in the $\mathcal{PT}$-symmetry preserving regime, with an energy constraint that the difference between the two eigenvalues of the $\mathcal{PT}$-symmetric Hamiltonian is real, and experimental attempts are only based on optical systems. 
Considering the close connection with no-cloning theorem~\cite{zurek-1982,bennett-2014}, security of quantum cryptographic protocols~\cite{bennett-1992,phoenix-1995,gisin-2002}, no-go theorems in quantum theory~\cite{pusey-2012,wei-2022}, and maximum mutual information~\cite{sasaki-1999}, it is of both fundamental importance and practical interest to extend the scope of $\mathcal{PT}$-symmetric QSD to wider parameter regimes and other platforms.

Here, we present the first experimental realization of $\mathcal{PT}$-symmetric QSD in a single trapped ion, and extend the scope to both the $\mathcal{PT}$-symmetry preserving (PTS) and broken (PTB) regimes. By employing a spontaneous decay process to bring in non-Hermiticity, we observe that two non-orthogonal states can become orthogonal in both the PTS and PTB regimes, such that a deterministic state discrimination can be achieved. We measure the orthogonal time by varying the dissipation parameter and the relative angle between the two candidate states, and map out the region of successful orthogonalization. Interestingly, we find that for a given pair of candidate states, there exists an optimal $\mathcal{PT}$-symmetric Hamiltonian giving the fastest speed of orthogonalization, i.e., a quantum brachistochrone for QSD. The optimal Hamiltonian may be in the PTS or PTB regime, depending on the states about to be discriminated. All results can be understood with a clear geometric picture, from which many analytic results can be extracted.

{\it {$\mathcal{PT}$-symmetric QSD}}.--
We consider a two-level $\mathcal{PT}$-symmetric Hamiltonian
\begin{eqnarray}
\label{Hpt}
\mathcal{H}_{\mathcal{PT}}
= i \Gamma \sigma_z + J \sigma_x,
\end{eqnarray}
where $\sigma_{i=x,z}$ are Pauli matrices. Without loss of generality, we assume that the two-level coupling strength $J$ and the dissipation rate $\Gamma$ are both real and positive. This Hamiltonian possesses $\mathcal{PT}$ symmetry with $[\mathcal{PT}, \mathcal{H}_{\mathcal{PT}}]=0$, with parity operator $\mathcal{P}=\sigma_x$ and time-reversal operator $\mathcal{T}$ being complex conjugate. By defining a dimensionless dissipation parameter $a \equiv \Gamma/J$, we can obtain a re-parameterized form as $\mathcal{H}_{\mathcal{PT}} = \omega {\pmb \sigma}\cdot {\bf n}$, where ${\pmb \sigma} =(\sigma_x,\sigma_y,\sigma_z)$, and ${\bf n} = (J,0,i\Gamma)/\omega$ with $\omega = J \sqrt{1-a^2} =\sqrt{J^2-\Gamma^2}$ representing a unit vector in spin space. Notice that $\omega$ is real and positive in the PTS regime of $a\in[0,1)$, and positively imaginary in the PTB regime of $a\in(1,+\infty)$. The two phases join at an exceptional point (EP) located at $a=1$~\cite{kato-1966}.

For any two states $\ket{\psi_{1,2}} $ about to be discriminated, one can always choose the spin coordinates accordingly to make the two states lie on the $y$-$z$ plane and be symmetric with respect to the $y$ axis. The two states are then expressed as $\ket{\psi_{1,2}} = ({\rm cos} \frac{\pi\mp2\theta}{4}, e^{i\phi}{\rm sin}\frac{\pi\mp2\theta}{4})^T$ in the complete basis $\{\ket{\uparrow_z}, \ket{\downarrow_z}\}$ and with the conventional inner product $\bra{\psi_1}\ket{\psi_2}={\rm cos}\theta$. Here, $\theta\in(0,\pi/2)$ is half the angle between $\ket{\psi_1}$ and $\ket{\psi_2}$ on the Bloch sphere, and $\phi$ is the relative phase. The evolution of the two states subjected to $\mathcal{H}_{\mathcal{PT}}$ can be obtained with the aid of the standard matrix identity $e^{i\omega {\pmb \sigma}\cdot{\bf n}} = {\rm cos}\omega\,\sigma_0 + i\,{\rm sin}\omega\,{\pmb \sigma}\cdot{\bf n}$, and expressed in spin coordinates $r_{i=1,2}(t)={\rm tr}[\ket{\psi_i(t)}\bra{\psi_i(t)}\sigma_r]$~\cite{nelson-2000} with $r=\{x,y,z\}$.
Under the condition of $\phi=(2n-1/2)\pi$ we can trace the state evolution by their corresponding coordinates.
%
%
The time evolution operator can be regarded as a rotation along the $x$ axis but with a time-dependent angular velocity, which is controlled by the parameter $a$. Since the evolution operator is not unitary, the locus of $\ket{\psi_{1,2}(t)}$ is a segment of a conical section satisfying
$\frac{[y_{1,2}(t) + D]^2}{A^2}+\frac{b^2}{|b|^2} \frac{z_{1,2}^2(t)}{B^2}=1$, 
%
%
with the coordinates of the center $O'=(0,-D,0)$ and $D=\frac{1-a\,{\rm cos}\theta}{|1-a\,{\rm cos}\theta|}\frac{b^2}{|b|^2}C$. Here, $C=a|1-a\,{\rm cos}\theta|/|b|^2$ is half focal length, and $A=|1-a\,{\rm cos}\theta|/|b|^2$ and $B = |1-a\,{\rm cos}\theta|/|b|$ are the semi-major and semi-minor axes, respectively.

\begin{figure}[t]
	\begin{center}
		\includegraphics[width=0.95\linewidth]{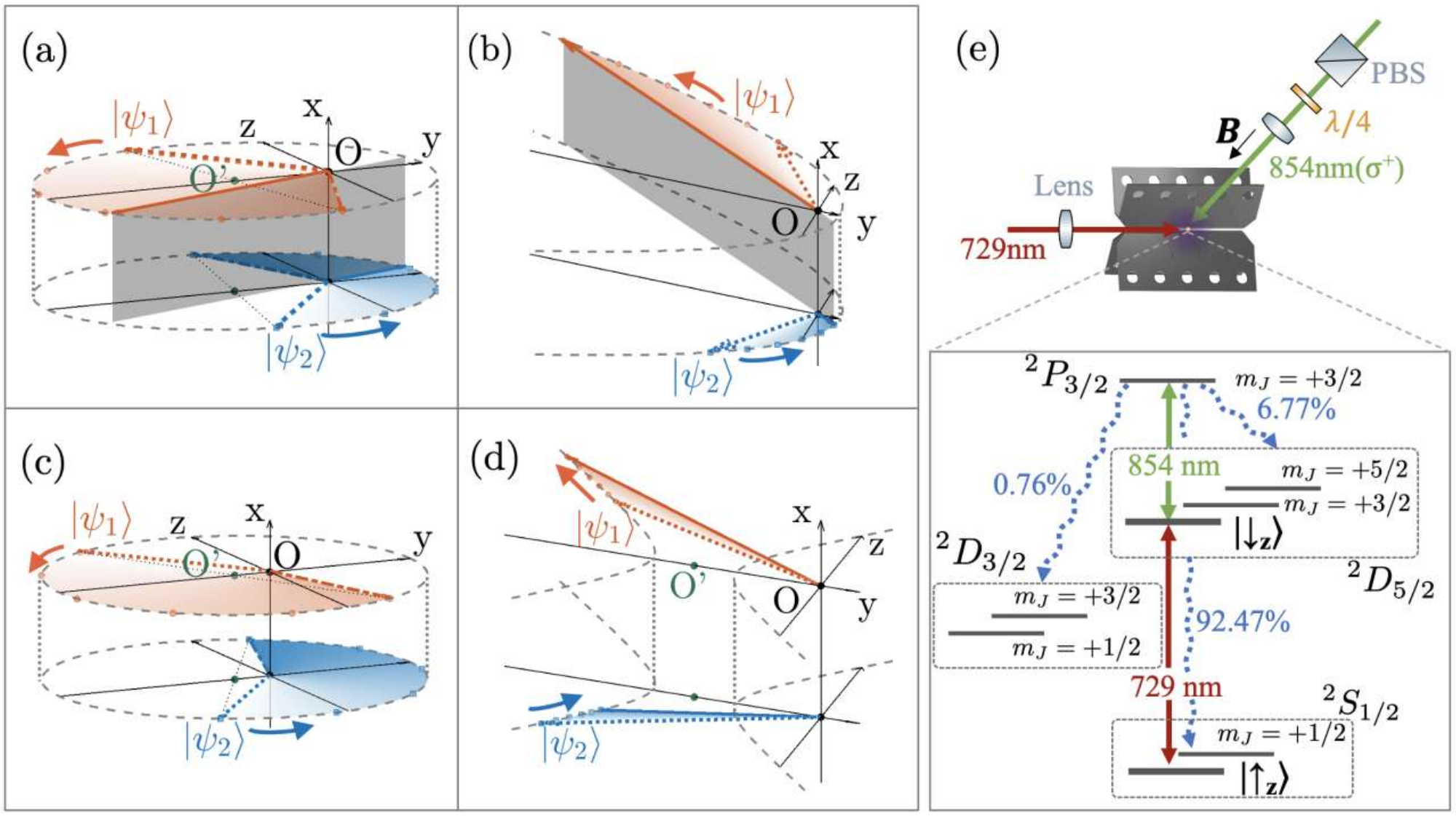}
	\end{center}
	\vspace{-8mm}
	\caption{
	Evolution of two initial states $\ket{\psi_1}$ and $\ket{\psi_2}$ on the Bloch sphere in (a), (c) PTS regime or (b), (d) PTB regime. The initial states are in the $y$-$z$ plane and symmetric about the $y$ axis. The two states evolve on the locus (gray dashed curves) of an ellipse in (a), (c) and a hyperbola in (b), (d).
In (a) and (b), the two states rotate counterclockwise and become antiparallel (orthogonal) at the positions denoted by arrows with solid body and highlighted by the shaded plane.
In (a) and (c), two states $\ket{\psi_{1,2}(T/2)}$ at a half period $t=T/2$ denoted by arrows with dot-dashed body are centra-symmetric with the initial state $\ket{\psi_{1,2}}$ about the ellipse center $O'$, and are located in the third and second quadrants in (a) for $a_l\le a<1$, but in the fourth and first quadrants in (c) for $a<a_l$. 
(e) Schematic experimental setup and level diagram of $^{40}{\rm Ca}^+$ ion. Detailed discussion can be found in Supplementary Materials~\cite{supp}.
	}
	\label{fig1}
\end{figure}

In the PTS regime with $a <1$, the locus is an ellipse with the right focus at the origin of spin space. In the PTB regime with $a >1$, the states are moving along the left sector of a hyperbola with the left focus at the origin when $1<a<1/{\rm cos}\theta$, or with the right focus at the origin when $a>1/{\rm cos}\theta$. We illustrate the evolution of Bloch state-vectors for the PTS and PTB regimes in Figs.~\ref{fig1}(a) and \ref{fig1}(b), respectively, where two non-orthogonal states can be successfully orthogonalized. 
For the PTS case shown in Fig.~\ref{fig1}(a), two initial states $\ket{\psi_1}$ and $\ket{\psi_2}$ denoted by red and blue arrows with dotted body rotate counterclockwise along an ellipse, and become orthogonal (antiparallel) at the positions labeled by red and blue arrows with solid body. For the PTB case shown in Fig.~\ref{fig1}(b), the two states rotate counterclockwise on the left sector of a hyperbola and can also become orthogonal. 

Another finding is that for a given pair of states separated by an angle $2 \theta$, there exist a lower bound of the dissipation parameter $a_l=(1-{\rm sin}\theta)/{\rm cos}\theta$ in the PTS regime and an upper bound $a_u=1/{\rm cos}\theta$. When $a<a_l$, the evolution is only slightly modified from a Hermitian system, the difference of rotation speed is too small to overcome the relative angle between the two states, such that they can never become orthogonal as depicted in Fig.~\ref{fig1}(c). For the upper bound in the PTB regime, remind that the memory of an initial state will continuously flow into the environment until completely dissipated, and the state finally reaches the eigenstate $\ket{\psi_+} = (i(a+|b|),1)^{\rm T}$ associated with the eigenvalue $E_+$ with a positive imaginary part~\cite{ding-2022}. When $a>a_u$, the asymptotic state $\ket{\psi_+}$ lies above the state $\ket{\psi_1}$ in the $y$-$z$ plane. During evolution, $\ket{\psi_{1,2}}$ rotate clockwise to approach the final state and can never become orthogonal as seen from Fig.~\ref{fig1}(d). 


One can also understand the bounding conditions from the geometry of evolution trajectories. For the PTS regime, when the states evolve a half period, they become centra-symmetric with the initial state about the ellipse center $O'$, as depicted by arrows with dot-dashed body in Figs.~\ref{fig1}(a) and \ref{fig1}(c). If $\ket{\psi_{1,2}(T/2)}$ are in the third and second quadrants as shown in Fig.~\ref{fig1}(a), they must be orthogonal at a certain time before $T/2$, since the angle from $\ket{\psi_1}$ to $\ket{\psi_2}$ is initially less than $\pi$ at $t=0$ and larger than $\pi$ at $t=T/2$. However, if $a$ is decreased below the critical value $a_l$, $\ket{\psi_{1,2}(T/2)}$ lie in the fourth and first quadrants with the angle from $\ket{\psi_1}$ to $\ket{\psi_2}$ being always smaller than $\pi$, as shown in Fig.~\ref{fig1}(c). Thus, the critical value of $a_l$ can be determined geometrically by setting the string connecting the two initial states $\ket{\psi_{1,2}(0)}$ passing through the left focal point of the ellipse. For the PTB regime, when $1<a<a_u$, the origin $O$ is located at the left focal point as depicted in Fig.~\ref{fig1}(b). While $\ket{\psi_1}$ is kept in the second quadrant, $\ket{\psi_2}$ will sweep around the origin counterclockwise through the fourth and first quadrants, such that they must be orthogonal at some point. However, if $a > a_u$, the origin moves to the right focal point, and both $\ket{\psi_1}$ and $\ket{\psi_2}$ are kept in the second and third quadrants, as seen from Fig.~\ref{fig1}(d). The critical $a_u$ thus corresponds to the case where the hyperbola become degenerate. 

{\it {Experiment realization and quantum brachistochrone}}.--
To connect with trapped ion systems, where state population is a natural observable, we invoke the time-dependent normalized population on basis $\ket{s}$ for state $\ket{\psi_{i=1,2}}$, defined as
\begin{eqnarray}
	{\color{black}
	\bar{P}_{i,s}^{\mathcal{PT}}(t) = \frac{P^{\mathcal{PT}}_{i,s}(t)}{P^{\mathcal{PT}}_{i,z_+}(t) + P^{\mathcal{PT}}_{i,z_-}(t)},}
\end{eqnarray}
where {\color{black} $P^{\mathcal{PT}}_{i,s}(t)=|\bra{s}e^{-i\mathcal{H}_{\mathcal{PT}}t}\ket{\psi_i}|^2$} is the corresponding unnormalized state population, and $\ket{s}$ is one of the eigenvectors of Pauli matrices $\sigma_y$ and $\sigma_z$, denoted by $\ket{y_{\pm}}\equiv\ket{\uparrow(\downarrow)_y}=(i,\mp 1)^{\rm T}/\sqrt{2}$ and $\ket{z_{\pm}}\equiv\ket{\uparrow(\downarrow)_z}$, respectively. 
Once the two states are orthogonal, the normalized populations on state $\ket{z_{\pm}}$ satisfy {\color{black} $\bar{P}^{\mathcal{PT}}_{1,z_+}(t)=\bar{P}^{\mathcal{PT}}_{2,z_-}(t)$ or $\bar{P}^{\mathcal{PT}}_{2,z_+}(t)=\bar{P}^{\mathcal{PT}}_{1,z_-}(t)$}. 
We also need to measure {\color{black} $\bar{P}^{\mathcal{PT}}_{1,y_+}(t)$ and $\bar{P}^{\mathcal{PT}}_{2,y_+}(t)$} to exclude the case where the final states are symmetric about the $y$ axis but not orthogonal.
This criterion also holds for a mapped purely dissipative Hamiltonian $\mathcal{H}_{\rm Diss}=\mathcal{H}_{\mathcal{PT}}-i\Gamma \sigma_0$, whose unnormalized population {\color{black} $P_{i,s}^{\rm Diss}(t) = e^{-2\Gamma t} P^{\mathcal{PT}}_{i,s}(t)$} has an exponentially decayed pre-factor.

We realize the two-level dissipative Hamiltonian $\mathcal{H}_{\rm Diss}$ by an eight-level non-Hermitian model of a single trapped $^{40}{\rm Ca}^+$ ion~\cite{chen-2021}, as illustrated in Fig.~\ref{fig1}(e). The two-level system is constructed by two Zeeman sublevels $\ket{\downarrow_z}\equiv\ket{^2D_{5/2},m_J=+1/2}$ and $\ket{\uparrow_z}\equiv\ket{^2S_{1/2},m_J=-1/2}$, and the transition is driven by a laser at wavelength 729 nm, with coupling strength $J$ measured by fitting the Rabi frequency. Another 854 nm beam is used to excite the ion from $\ket{\downarrow_z}$ to the excited $\ket{^{2}P_{3/2},m_J=+3/2}$ state with coupling strength $J_c$, which decays quickly to $\ket{^2D_{5/2},m_J=+1/2,+3/2,+5/2}$, $\ket{^2S_{1/2},m_J=+1/2}$ and $\ket{^2D_{3/2},m_J=+1/2,+3/2}$ with transition rates $\Gamma_1$, $\Gamma_2$ and $\Gamma_3$, respectively. {\color{black} The corresponding branching ratios are experimentally reported as $5.87\%$, $93.5\%$ and $0.63\%$~\cite{gao-2019}. This process induces a tunable loss on the $\ket{\downarrow_z}$ state with an effective dissipation rate $\Gamma= J_c^2(\Gamma_0-\gamma_1)/\Gamma_0^2$~\cite{ding-2021,supp}, where the total transition rate $\Gamma_0=\Gamma_1+\Gamma_2+\Gamma_3$ of the $^2P_{3/2}$ state is obtained by its lifetime $\tau=6.639$ ns~\cite{meir-2020}, and the transition rate from $\ket{^2P_{3/2},m_J=+3/2}$ to $\ket{\downarrow_z}$ state is $\gamma_1=\Gamma_1/15$}. We also calibrate the effective dissipation rate from a fitting of exponential decay by preparing the initial state on state $\ket{\downarrow_z}$ and turning on the 854 nm dissipative beam only. The population on state $\ket{\uparrow_z}$ is measured by the standard fluorescence counting rate threshold method~\cite{sage-2019}, while the population on state $\ket{\downarrow_z}$ is calculated by the measurement of population on $^2S_{1/2}$ manifold. 
As for the population on state $\ket{y_+}$, we first operate $\sigma_x$ on state $\ket{\psi_i}$ for a period of $J t=\pi/4$, which corresponds to an evolution operator $\mathcal{U}_+=e^{-i {\pi} \sigma_x/4}$, then detect on the basis $\ket{\uparrow_z}$.

\begin{figure}[!t]
	\begin{center}
		\includegraphics[width=0.9\linewidth]{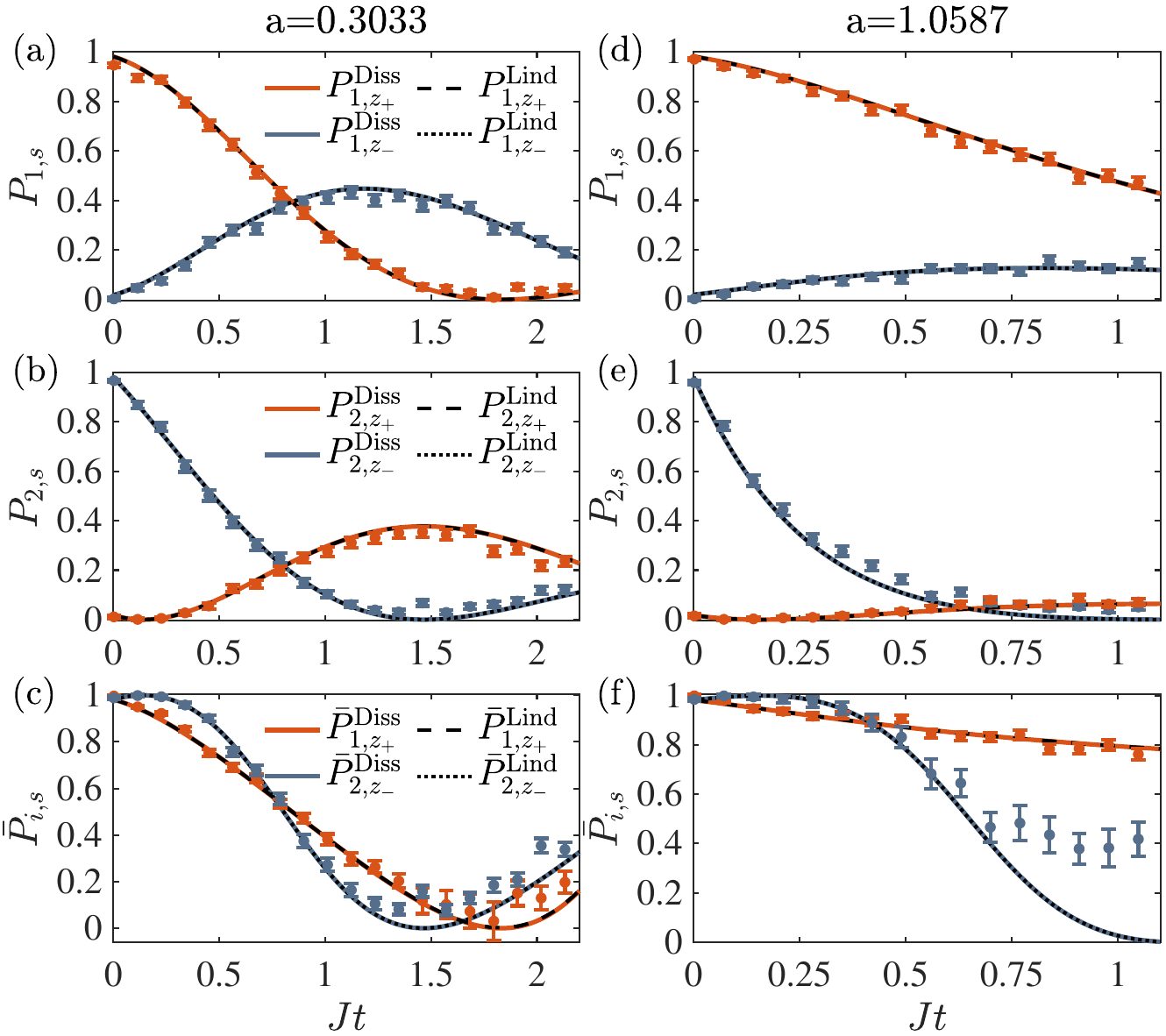}
	\end{center}
	\vspace{-8mm}
	\caption{{\color{black}(a) Time evolution of {\color{black} $P_{1,z_+}^{\rm Diss}(t)$} (red solid line) and {\color{black} $P_{1,z_-}^{\rm Diss}(t)$} (blue solid line) governed by the dissipative Hamiltonian with parameter $a=\Gamma/J=0.3033$, where the initial state is $\ket{\psi_1}$ with $\theta=1.3$. Populations of {\color{black} $P_{1,z_+}^{\rm Lind}$} (black dashed line) and {\color{black} $P_{1,z_-}^{\rm Lind}$} (black dotted line) given by the Lindblad equation are shown for comparison. The same results for initial state $\ket{\psi_2(\theta=1.3)}$ are shown in (b).
(c) The normalized populations of {\color{black} $\bar{P}_{1,z_+}^{\rm Diss}$} (red solid line), {\color{black} $\bar{P}_{2,z_-}^{\rm Diss}$} (blue solid line), {\color{black} $\bar{P}_{1,z_+}^{\rm Lind}$} (black dashed line) and {\color{black} $\bar{P}_{2,z_-}^{\rm Lind}$} (black dotted line). Panels (d)-(f) are the counterparts of panels (a)-(c), with the effective dissipation $a=1.0587$. In panels (a)-(b) and (d)-(e), red and blue dots are the corresponding experimental data averaged over 500 rounds of measurement with error bars obtained by the standard deviation. The data shown in panels (c) and (f) are obtained from (a)-(b) and (d)-(e) from normalization, respectively.}
}
	\label{fig2}
\end{figure}

In Fig.~\ref{fig2}, we show the time evolution of unnormalized population {\color{black} $P_{1,z_+}^{\rm Diss}$} on state $\ket{z_+}$ and {\color{black} $P_{1,z_-}^{\rm Diss}$} on $\ket{z_-}$ with initial state $\ket{\psi_1(\theta=1.3)}$ in panel (a). Parameters of $J$ and $\Gamma$ are chosen to give $a=\Gamma/J=0.3033$, i.e., in the PTS regime. The same quantities are plotted in panel (b) but with initial state $\ket{\psi_2(\theta=1.3)}$. We also present the results of {\color{black} $P_{i,z_+}^{\rm Lind}(t)$ and $P_{i,z_-}^{\rm Lind}(t)$} with the same initial states $\ket{\psi_{i=1,2}(\theta=1.3)}$, which are obtained by numerically evolving the Lindblad equation with 
{\color{black} $\Gamma_1/J= 157.7930$, $\Gamma_2/J=2513.3976$, $\Gamma_3/J=16.9352$ and $J_c/J=28.6109$, }
corresponding to the same effective dissipation $a=0.3033$. The normalized populations {\color{black} $\bar{P}_{1,z_+}^{\rm Diss/Lind}$ and $\bar{P}_{2,z_-}^{\rm Diss/Lind}$} are presented in Fig.~\ref{fig2}(c).
These results show good agreement with the experimental data denoted by dots, obtained with compatible parameters of 
{\color{black} $\Gamma_1 = 2\pi\times1.4072$ MHz, $\Gamma_2 = 2\pi\times22.4145$ MHz, $\Gamma_3 =2\pi\times 0.1510$ MHz~\cite{gao-2019,meir-2020},} and a Rabi frequency $J = 0.0560$ MHz. 
Besides, we find the deviation between the two-level $\mathcal{PT}$-symmetric system and Lindblad equation is quite small within the time scale shown in Fig.~\ref{fig2}.
The experimental orthogonal time $Jt_{\rm orth}^{\rm Exp}=0.7420$ can be extracted by fitting experimental data, which is close to the intersection 
{\color{black} $Jt_{\rm orth}^{\rm Lind}=0.7564$} between {\color{black} $\bar{P}_{1,z_+}^{\rm Lind}$} (black dashed) and {\color{black} $\bar{P}_{2,z_-}^{\rm Lind}$} (black dotted), and the orthogonal time $Jt_{\rm orth}^{\rm Diss}=0.7566$ predicted by the two-level dissipative Hamiltonian. 

The results for the PTB regime are shown in Figs.~\ref{fig2}(d)-\ref{fig2}(e), with 
{\color{black} $J_c=2.3627$ MHz} and $J = 0.0349$ MHz, corresponding to a larger dissipation parameter $a= 1.0587$. In such case, {\color{black} $\bar{P}_{1,z_+}^{\rm Lind}$ and $\bar{P}_{2,z_-}^{\rm Lind}$} cross at 
{\color{black} $Jt_{\rm orth}^{\rm Lind}=0.4179$}, and the orthogonal time determined by $\mathcal{H}_{\rm Diss}$ is $Jt_{\rm orth}^{\rm Diss}=0.4183$, both showing good consistency with the experimental outcome $Jt_{\rm orth}^{\rm Exp}=0.4219$. Thus, we demonstrate a successful $\mathcal{PT}$-symmetric QSD in both PTS and PTB regimes. 

\begin{figure}[t]
	\begin{center}
		\includegraphics[width=0.9\linewidth]{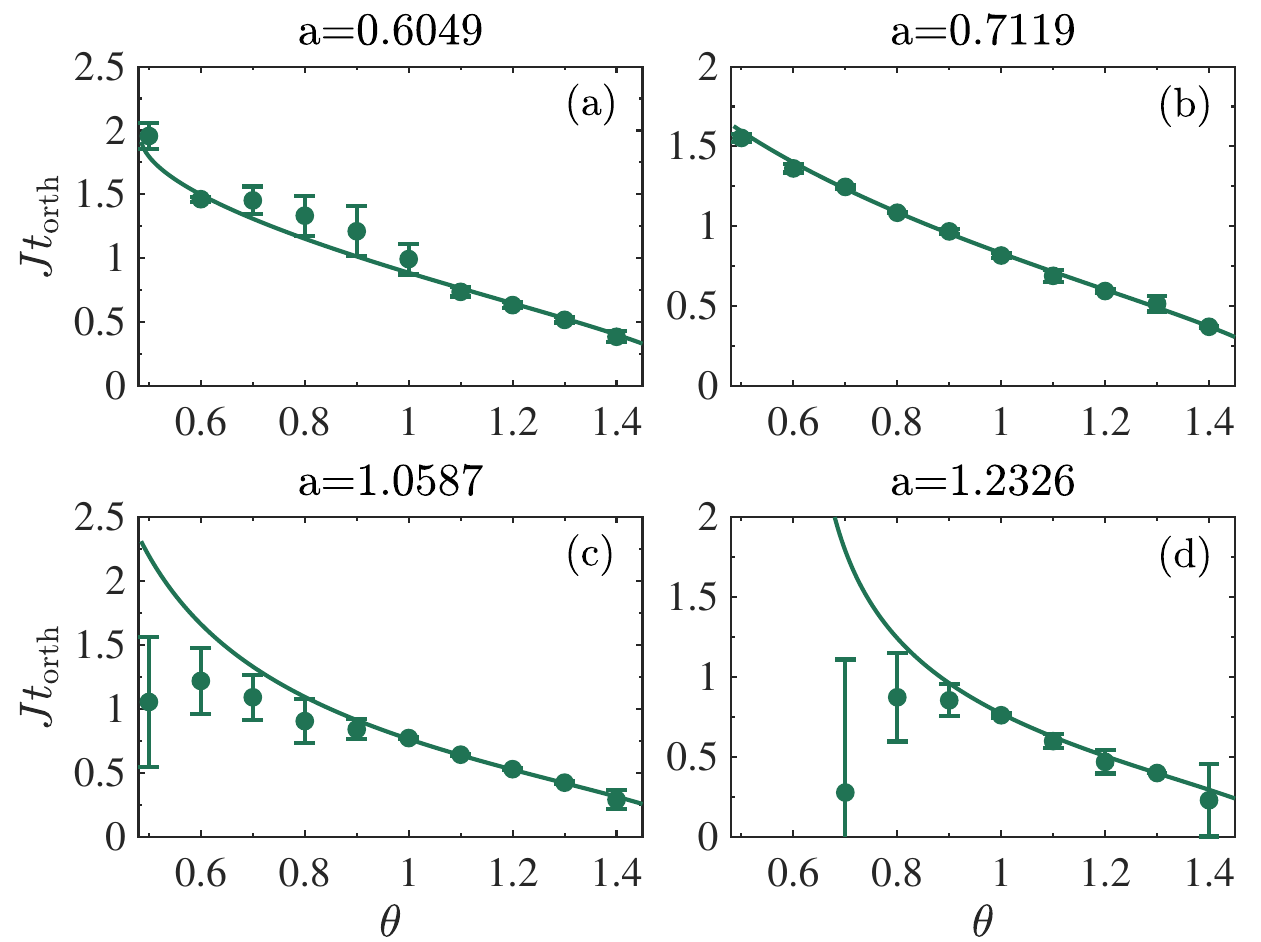}
	\end{center}
	\vspace{-8mm}
	\caption{The orthogonal time $t_{\rm orth}$ as a function of relative angle $\theta$. The solid lines are theoretical results of a $\mathcal{PT}$-symmetric Hamiltonian. The experimental data marked by dots are obtained by fitting the evolution of state populations. 
	}
	\label{fig3}
\end{figure}

In Fig.~\ref{fig3}, we present the orthogonal time as a function of the relative angle $\theta$ between initial states, with different dissipation parameters $a=0.6049, 0.7119, 1.0587, 1.2326$. Results from $\mathcal{PT}$-symmetric Hamiltonian (lines) and experiment (dots) are shown for comparison. The dimensionless orthogonal time $Jt_{\rm orth}$ is a decreasing function of $\theta$, as one would naturally expect. The experimental results agree with the two-level predictions in the PTS regime, but deviate apparently in the PTB regime as $\theta$ becomes smaller and the parameter $a$ approaches the upper bound $a_u = 1/\cos\theta$. The reasons for the notable deviation are analyzed in Supplementary Material~\cite{supp}. 

\begin{figure}[!t]
	\begin{center}
		\includegraphics[width=1.0\linewidth]{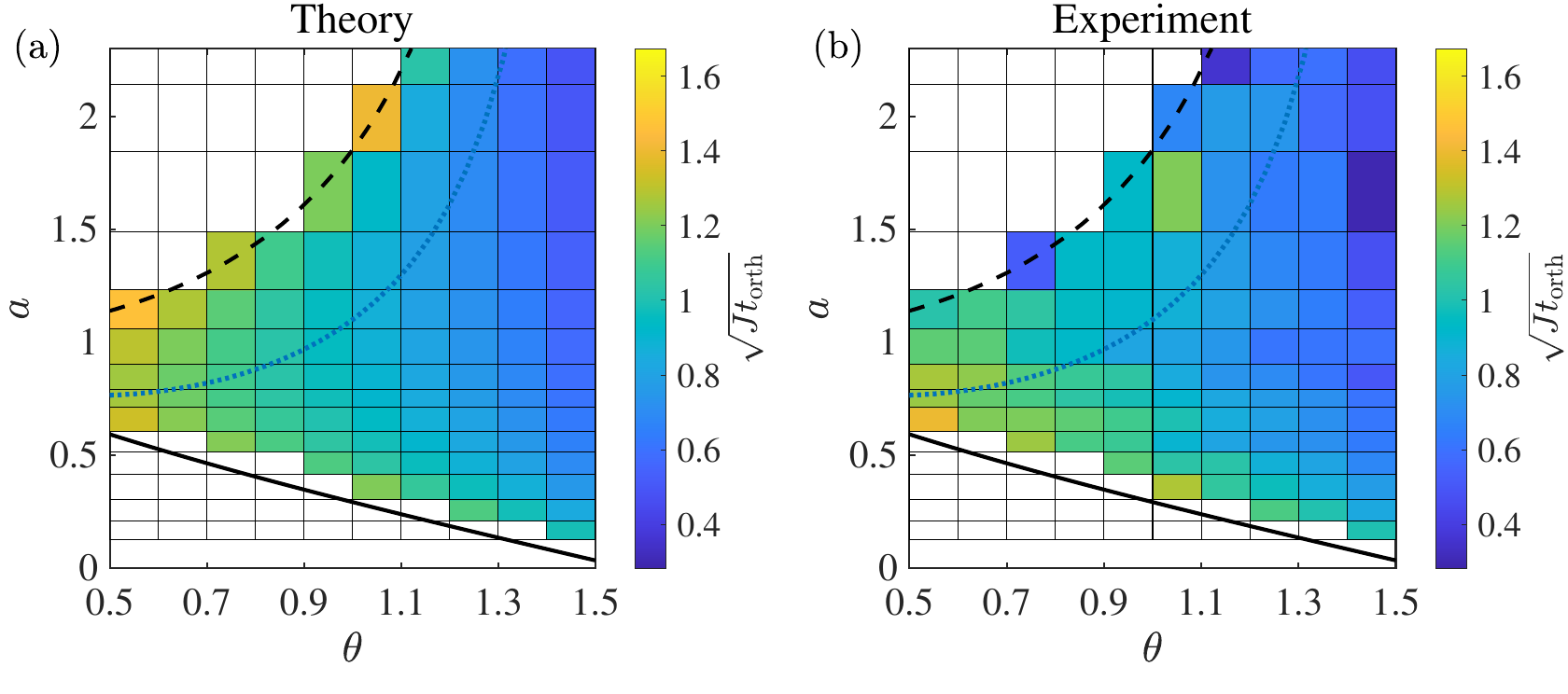}
	\end{center}
	\vspace{-8mm}
	\caption{The valid region of $\mathcal{PT}$-symmetric QSD is enclosed by the upper bound (black dashed line) and lower bound (black solid line), within which the orthogonal time is depicted by false-color. The blue dotted line represents the fastest possible time of orthogonalization. The parameters for $a$ and $\theta$ are referred to the left-bottom corner of each colored brick cell, and $\sqrt{Jt_{\rm orth}}$ is used to enhance the color distinguishability.
	 }
	\label{fig4}
\end{figure}

Finally, we measure the orthogonal time by varying the initial states and effective dissipation parameter. The theoretical prediction based on a two-level $\mathcal{PT}$-symmetric Hamiltonian are shown in Fig.~\ref{fig4}(a). The black solid and dashed lines depict the lower and upper bounds of $a$, respectively. The enclosed area is the region for a successful $\mathcal{PT}$-symmetric QSD, and the orthogonal time is shown in false-color. One can see clearly that the lower (upper) bound of $a$ lies in the PTS (PTB) regime, and decreases (increases) with $\theta$, such that the range of $a$ gets expanded. Intuitively, an arbitrary $\mathcal{PT}$-symmetric Hamiltonian can orthogonalize two states which are initially orthogonal with $\theta=\pi/2$, while it is impossible to do so for two identical states with $\theta=0$. 
Interestingly, for a given pair of candidate states, there is an optimal choice of $a$ such that the time required for orthogonalization is minimal, i.e., a quantum brachistochrone of $\mathcal{PT}$-symmetric QSD. The optimized parameter $a(t_{\rm opt})$ may be in the PTS or PTB regimes, as represented by blue dotted line in Fig.~\ref{fig4}(a). 
The experimental results are presented for comparison in Fig.~\ref{fig4}(b), and show good agreement in the PTS regime and shallow PTB regime. A significant deviation is observed when the dissipation is strong and the initial states are approaching the upper bound, which is analyzed in Supplementary Material~\cite{supp}. 

{\it {Summary}.}--
We experimentally demonstrate unambiguous quantum state discrimination (QSD) of two non-orthogonal two-level states in a single $^{40}{\rm Ca}^+$ ion with the aid of a $\mathcal{PT}$-symmetric Hamiltonian. For the first time, we realize a successful $\mathcal{PT}$-symmetric QSD in both the $\mathcal{PT}$-symmetry preserving and broken regimes, that is, the two states can become orthogonal upon evolution and can be deterministically discriminated with a probability of null measurement. We measure the time of orthogonalization by varying the dissipation parameter and the relative angle between the two candidate states, and map out the region for a successful orthogonalization in parameter space. For a given pair of candidate states, we find the optimal dissipation parameter to realize the fastest orthogonalization, i.e., a quantum brachistochrone of $\mathcal{PT}$-symmetric QSD. These results can be understood within a clear geometric picture, from which many analytic results can be obtained. Our work demonstrates a practical application of non-Hermitian physics in quantum tasks.

{\it {Acknowledgements.}}--
We thank Pingxing Chen, Yiheng Lin, and Chunwang Wu for fruitful discussion. This work is supported by the National Key R\&D Program of China (Grants No.~2022YFA1405301), and the National Natural Science Foundation of China (Grants No.~12074427, No.~12074428, and No.~92265208).

{$^{\#}$ C. Z. and T. S. contributed equally to this work.}

\end{document}